\documentclass[useAMS,usenatbib]{mn2e} 
\usepackage{graphicx}
\usepackage{epsfig}
\usepackage{amsmath}
\usepackage{color}

\newcommand{\Tigm}{\mathcal{T}_{\rm IGM}}

\newcommand{\avenf}{\bar{x}_{\rm HI}}

\newcommand{\lya}{Ly$\alpha$}

\newcommand{\hs}{\hspace{1mm}}

\newcommand{\Msun}{M_\odot}

\newcommand\lsim{\mathrel{\rlap{\lower4pt\hbox{\hskip1pt$\sim$}}
        \raise1pt\hbox{$<$}}}
\newcommand\gsim{\mathrel{\rlap{\lower4pt\hbox{\hskip1pt$\sim$}}
        \raise1pt\hbox{$>$}}}
\newcommand\ion[2]{#1$\;${\scshape{#2}}}

\newcommand{\apj}{ApJ}
\newcommand{\apjl}{ApJ}
\newcommand{\apjs}{ApJS}
\newcommand{\aap}{A\&A}
\newcommand{\aj}{AJ}
\newcommand{\mnras}{MNRAS}
\newcommand{\pasj}{PASJ}
\newcommand{\physrep}{Physics Reports}

\newcommand{\nat}{Nature}

\def\lsim{~\rlap{$<$}{\lower 1.0ex\hbox{$\sim$}}}

\def\gsim{~\rlap{$>$}{\lower 1.0ex\hbox{$\sim$}}}

\title[The Detectability of LAEs during the EoR]{The Detectability of Ly$\alpha$ Emission from
  Galaxies during the Epoch of Reionization}

\author[Dijkstra, Mesinger \& Wyithe]{Mark
  Dijkstra$^{1}$\thanks{E-mail: dijkstra@mpa-garching.mpg.de}, Andrei
  Mesinger$^{2}$\thanks{Hubble Fellow; E-mail:
    mesinger@astro.princeton.edu} and J. Stuart
  B. Wyithe$^{3}$\\ $^{1}$Max Planck Institute fur Astrophysik, Karl-Schwarzschild-Str. 1, 85741 Garching, Germany\\ 
$^{2}$Department of Astrophysical Sciences, Princeton University,
  Princeton, NJ 08544, USA \\ $^{3}$School of Physics, University of
  Melbourne, Parkville, Victoria, 3010, Australia}

\begin{document}

\date{\today} \pagerange{\pageref{firstpage}--\pageref{lastpage}}
\pubyear{2009}

\voffset-.6in

\maketitle
\label{firstpage}

\begin{abstract}
We study the visibility of the Ly$\alpha$ emission line during the
Epoch of Reionization (EoR). Combining galactic outflow models with
large-scale semi-numeric simulations of reionization, we quantify the
probability distribution function (PDF) of the fraction of
\lya\ photons transmitted through the intergalactic medium
(IGM), $\mathcal{T}_{\rm IGM}$. Our study focusses on galaxies populating dark
matter halos with masses of $M_{\rm halo}=10^{10}M_{\odot}$ at
$z=8.6$, which is inspired by the recent reported discovery of a
galaxy at $z=8.6$ with strong Ly$\alpha$ line emission.
 For reasonable assumptions,  we find that the combination of winds and reionization morphology results in $\mathcal{T}_{\rm IGM}\gsim 10 \%$ [50\%], for the majority of galaxies, even when the Universe is $\sim 80 \%$
  [60\%] neutral by volume.  Thus, the observed strong Ly$\alpha$
  emission from the reported $z=8.6$ galaxy is consistent with a
  highly neutral IGM, and cannot be used to place statistically significant constraints on the volume averaged neutral fraction of hydrogen in the IGM.
 We also investigate the implications of the recent 
  tentative evidence for a observed decrease in the `Lyman Alpha Emitter fraction' among drop-out galaxies between $z=6$ and $z=7$. If confirmed, we show that a rapid evolution in $\avenf$ will be required to explain this observation via the effects of reionization.
 \end{abstract}

 
\section{Introduction}
\label{sec:intro}

The Wide Field Camera 3 on board of the {\it Hubble Space Telescope} has enhanced our ability to observe galaxies at redshifts great than
six, so-far obtaining $\sim 100$ likely candidate galaxies at $z=$7, 8 \citep[e.g.][]{Bouwens10,Bunker10,Finkelstein10,Yan10}. The {\it James
  Webb Space Telescope} is expected to probe galaxies a few magnitudes deeper, and also to spectroscopically confirm the redshifts
of the existing candidates. One of the key predicted properties of young, metal poor galaxies in
the high-redshift Universe are prominent nebular emission lines,
dominated by hydrogen Ly$\alpha$ ($\lambda=1216$\hs\AA, see
e.g. Johnson et al. 2009, Pawlik et al. 2010). The first generation of galaxies are likely
to have been strong Ly$\alpha$ emitters, with equivalent widths
possibly as high as EW$\sim 1500$ \AA \hs\citep[][also see Partridge
  \& Peebles 1967]{S02,S03,J09b}.

During the epoch of reionization (EoR), the Ly$\alpha$ emission line
may be difficult to observe, due to the large opacity of the
intervening neutral intergalactic medium: for example, a source needs
to be embedded in a $\gsim$1 Mpc HII region to allow Ly$\alpha$
photons to redshift far away from the line center before they reach
the IGM \citep[e.g.,][]{M98,Cen00}. For the galaxy to generate such a
large HII region, its ionizing luminosity would have to be
unphysically large (unless there is a bright quasar associated with
the galaxy, see Cen \& Haiman 2000). However, several effects  have
been shown to boost the detectability of the Ly$\alpha$ flux: ({\it
  i}) source clustering, which boosts the sizes of HII regions
\citep{Fur04,Mes2,McQuinn07,Iliev08} ({\it ii}) the patchiness of
reionization, which may give rise to significant fluctuations in the
IGM opacity between different sightlines, as well as a steeper
absorption profile \citep[e.g.,][]{Mes1,McQuinn08}, and ({\it iii})
radiative transfer effects through outflows of interstellar (ISM)
\ion{H}{I} gas, which can impart a redshift\footnote{The peculiar velocity of a galaxy can also redshift Ly$\alpha$ photons away from
  resonance before they escape into the surrounding intergalactic
  medium (Cen et al. 2005). However, this redshift is typically
  significantly smaller than the redshift imparted by galactic
  outflows (see Dijkstra \& Wyithe 2010).} to the Ly$\alpha$ photons
before they emerge from galaxies \citep[][but also see Barnes et al. 2011]{Santos04,DW10}.

\cite{Lehnert} recently reported a detection of strong Ly$\alpha$ line
emission from a $Y_{105}$ drop-out galaxy in Wide Field Camera 3
observations of the Hubble Ultra Deep Field. The Ly$\alpha$ line
implies that the galaxy is at $z\sim 8.56$, which is the highest
redshift of any spectroscopically-confirmed object to
date.
 Interestingly, when taken
at face value, the observed Ly$\alpha$ line is strong, with an
observed equivalent width (EW) of $\sim 100$ \AA\hs (see
\S~\ref{sec:discuss}). Motivated by this observation, we study the
visibility of the Ly$\alpha$ emission line during the EoR,
and  compute the total fraction of emitted Ly$\alpha$ photons that the IGM
transmits directly to the observer, $\mathcal{T}_{\rm IGM}$.
 We simultaneously include the inhomogeneous large-scale reionization
morphology, peculiar velocity offsets of the galaxies and IGM, and
radiative transfer through the galactic outflows that is calibrated by
observations of Lyman Alpha Emitters (LAEs) at $z<6$ \citep[][]{V06,V08,V10}. This, in combination with the fact that we compute the full $\mathcal{T}_{\rm IGM}$-PDF, clearly distinguishes our analysis from previous work. 
  
Our models do not include dust, which at the redshifts of interest
($z=7-9$) is likely a good approximation
\citep[][]{Stanway05,Bouwens10b,Hayes10,Blanc10}. The dust
opacity to Ly$\alpha$ photons inside galaxies is expected to increase towards lower
redshift, as the cumulative dust content of the Universe increased
with cosmic time. This expected evolution has a different sign than
the IGM opacity which decreases towards lower redshift. Thus, our
discussion regarding the redshift evolution of the IGM opacity is
likely conservative. 

The outline of this paper is as follows: in \S~\ref{sec:model}, we
describe our models for galactic outflows (\S \ref{sec:winds}) and IGM
opacity (\S \ref{sec:IGM}). In \S~\ref{sec:result} we present the
corresponding \lya\ transmission fractions.  Within this context, we
interpret the observations of Ly$\alpha$ emitting galaxies at $z>6$ in
\S~\ref{sec:discuss}. We compare our results with previous work in \S~\ref{sec:prevwork}.
Finally, we present our conclusions in
\S~\ref{sec:conc}.
 We adopt the background cosmological parameters ($\Omega_\Lambda$,
 $\Omega_{\rm M}$, $\Omega_b$, $n$, $\sigma_8$, $H_0$) = (0.72, 0.28,
 0.046, 0.96, 0.82, 70 km s$^{-1}$ Mpc$^{-1}$), matching the
 five--year results of the {\it WMAP} satellite
 \citep{Komatsu08}. Unless stated otherwise, we quote all quantities
 in comoving units.

\section{The Model}
\label{sec:model}

As mentioned above, our model has two components: (i) the intrinsic
\lya\ line, which has been processed by \lya\ scattering through galactic outflows, and (ii) the IGM opacity, which further processes
the line through both resonant and damping wing absorption.  We
describe these below in turn.

\subsection{Modeling Galactic Outflows}
\label{sec:winds}

We employ the wind models described in Dijkstra \& Wyithe (2010). In
these models the outflow is represented by a spherically symmetric
shell of HI gas that has a column density $N_{\rm HI}$, and outflow
velocity $v_{\rm wind}$. The shell has a thickness of 0.1 kpc and a
radius of 1.0 kpc (proper, also see Ahn et al. 2003). The outflow surrounds a central Ly$\alpha$ source which emits photons at line center. Since we focus solely on the total {\it fraction} of Ly$\alpha$ photons transmitted to the observer, the total Ly$\alpha$ luminosity of the source is irrelevant. A Monte-Carlo code
\citep[described in][]{D06} accurately follows the propagation of
Ly$\alpha$ photons through the optically thick outflow. These models
are very similar to the models that reproduce observed Ly$\alpha$ line
profiles at $z=3-5$ \citep[][see Dijkstra \& Wyithe 2010 for a
  discussion of caveats etc.]{V06,V08}. In this paper, we focus on
wind models with $N_{\rm HI}=10^{20}$ cm$^{-2}$ and $N_{\rm
  HI}=10^{21}$, and wind velocities of $v_{\rm wind}=25$ km s$^{-1}$
and $v_{\rm wind}=200$ km s$^{-1}$. For comparison, Verhamme et al
(2008) reproduced observed Ly$\alpha$ line shapes for LAEs at $z<5$
with 25 km/s$\leq v_{\rm wind} \leq$ 400 km/s, and that $2\times
10^{19}$ cm$^{-2} \leq N_{\rm HI} \leq 7 \times 10^{20}$
cm$^{-2}$. Our fiducial model ($N_{\rm HI}=10^{20}$ cm$^{-2}$, $v_{\rm
  wind}=200$ km s$^{-1}$) lies in the middle of this range. 
  
The impact of galactic winds on the Ly$\alpha$ radiation field depends on the covering factor of the galactic outflow: for low covering factors only a small fraction of the emitted Ly$\alpha$ photons will be Doppler boosted to frequencies where the IGM opacity is reduced. The issue of the wind covering factor is therefore related to the overall impact of galactic winds on the Ly$\alpha$ radiation field. Existing observations (which extend out to $z\sim 6$) indicate that winds play an important role in the scattering process, implying a large covering factor of the outflowing scattering material (we discuss this further in \S~\ref{sec:prevwork}). Of course, it remains a possibility that winds properties, such as their covering factor, were different at $z>6$. This would affect the impact of the (especially ionized) IGM on Ly$\alpha$ emission lines, and hence the observed redshift evolution of LAEs at these redshifts (see \S~\ref{sec:z7}).

\subsection{Modeling the IGM Opacity}
\label{sec:IGM}

We use the publicly-available, semi-numerical code
DexM\footnote{http://www.astro.princeton.edu/$\sim$mesinger/Sim.html}
to generate evolved density, velocity, halo, and ionization  fields at
$z=$ 8.56. This code and detailed tests are presented in Mesinger \&
Furlanetto (2007), Mesinger et al. (2010) and Zahn et al. (2010), to
which we refer the reader for details.  Here we briefly summarize our
simulation.

Our simulation box is $L=$ 250 Mpc on a side, with the final density,
peculiar velocity, and ionization fields having grid cell sizes of
0.56 Mpc.  Halos are filtered out of the 1800$^3$ linear density field
using the excursion-set formalism, and then mapped to Eulerian coordinates
at $z=8.56$ through perturbation theory (Zeldovich 1970).
Perturbation theory is also used to generate the evolved density and
peculiar velocity fields. Corresponding ionization fields are created
according to the excursion-set prescription described in Mesinger \&
Furlanetto (2007), with the modification from Zahn et al. (2010) to
account for partially ionized cells.  This prescription compares the
number of ionizing photons produced in a region of a given scale to
the number of neutral hydrogen atoms inside that region. We generate a
suite of ionization fields at various values of $\avenf$ by varying
the ionization efficiency of sources assumed to be hosted by
atomically-cooled halos, with masses of $M_{\rm halo} > 10^8 M_\odot$.  All of
these fields have been extensively tested against hydrodynamical
cosmological simulations, and good agreement was found well past the
linear regime (Mesinger \& Furlanetto 2007; Zahn et al. 2010; Mesinger
et al. 2010).

We then extract $\sim10^4$ line-of-sights (LOSs) centered on halos in the mass range,
$10^{10}\Msun < M_{\rm halo} < 3\times10^{10}\Msun$.  This choice of host halo
masses is motivated by the UV derived star formation rate (SFR) of 2-4
$\Msun$ yr$^{-1}$ \citep{Lehnert}, which corresponds to a halo mass of
$\sim10^{10} \Msun$ in the cosmological hydrodynamic simulations of
Trac \& Cen (2007)\footnote{Our results are not sensitive to
  uncertainties of a factor of few in the host halo mass, because the
  halo bias evolves relatively slowly with mass in this range
  (Mesinger \& Furlanetto 2008a; McQuinn et al. 2008; Mesinger \&
  Furlanetto 2008b).}. Opacities at wavelengths surrounding the
\lya\ line are computed for each LOS, integrating underneath a Voigt
absorption profile (e.g. Rybicki \& Lightman 1979), and including
contribution from both the ionized and neutral IGM, out to distances
of 200 Mpc away from the source.  Inside HII regions, a residual HI
fraction is computed assuming ionization equilibrium with an
ionization rate of $\Gamma_{\rm HII}=0.5 \times 10^{-12}$ s$^{-1}$, in
rough agreement with estimates obtained from the $z\sim$5--6
\lya\ forest (e.g., Fan et al. 2006; Bolton \& Haehnelt
2007)\footnote{We emphasize that there is currently no evidence that
  reionization has completed at these redshifts (Lidz et al. 2007;
  Mesinger 2010; McGreer et al. (2011).  If the
  observed quasar spectra go through regions of pre-overlap neutral
  gas, the inferred value of a homogeneous $\Gamma_{\rm tot}$ would
  include contributions from the neutral IGM ($\Gamma\sim0$).
  Therefore, the derived values of $\Gamma_{\rm tot}$ (from, e.g., Fan et al. 2006; Bolton \& Haehnelt 2007) can be treated
  as lower limits for the ionization rate {\it inside the ionized
    component of the IGM}, $\Gamma_{\rm HII}$.}. We include the
peculiar velocities of both the source halo and the absorbing gas, which can be very important \citep[e.g.][]{IGM,Iliev08,Laursen11}.

\begin{figure}
\vbox{\centerline{\epsfig{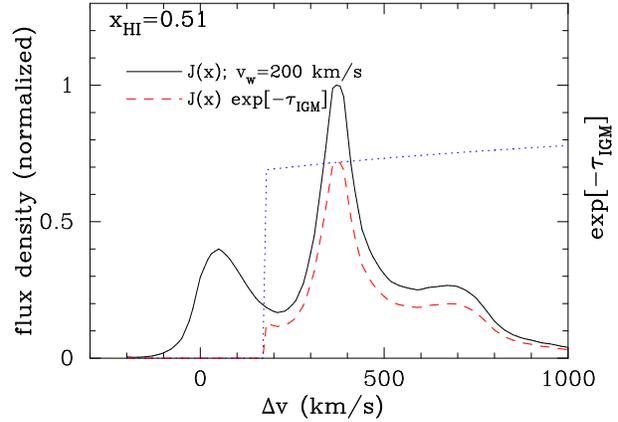}}}
\caption[]{This Figure shows an example of a Ly$\alpha$ line
  profile. The {\it solid line} shows the Ly$\alpha$ spectrum--in units
  of the peak flux density--that emerges from the galaxy after the
  photons have scattered through the HI outflow. The {\it blue dotted
    line} shows the IGM opacity, $\exp[-\tau_{\rm IGM}(\nu)]$ as a
  function of velocity offset $\Delta v$ for $\avenf=0.51$. The {\it
    red dashed line} shows the spectrum of photons after processing
  the flux through the IGM, i.e. $J(\nu)\exp[-\tau_{\rm IGM}(\nu)]$. This
  Figure nicely shows the impact of resonant scattering in the ionized
  IGM at $\Delta v \lsim 170$ km s$^{-1}$, and the impact of the
  damping wing optical depth at $\Delta v \gsim 170$ km s$^{-1}$. The
  latter varies only weakly with frequency. }
\label{fig:spec}
\end{figure}
\section{Results}
\label{sec:result}

To quantify the detectability of the \lya\ line, we compute for each
LOS the total fraction of emitted Ly$\alpha$ photons that the IGM
transmits directly to the observer, $\mathcal{T}_{\rm IGM}$, as\footnote{Photons
  that are scattered in the neutral IGM produce diffuse Ly$\alpha$ halos
  around individual sources \citep[e.g.][]{ZZ2}. This
emission is several
  orders of magnitude fainter than the detection threshold of the
  deepest observations to date (see Dijkstra \& Wyithe
  2010). Ly$\alpha$ radiation that is resonantly scattered in close
  proximity ($\lsim 10$ kpc) to the galaxy can give rise to a brighter
  Ly$\alpha$ halos \citep[][]{ZZ1,ZZ2}. However, winds reduce the brightness of
  these halos \citep{DW10}. In any case, by ignoring this resonantly
  scattered component, we underestimate the true Ly$\alpha$ flux that
  can be detected from galaxies during the EoR, which only renders our
  results conservative.}
\begin{equation}
\mathcal{T}_{\rm IGM}=\int_{\nu_{\rm min}}^{\nu_{\rm max}} d\nu
J(\nu)\exp[-\tau_{\rm IGM}(\nu)],
\label{eq:TIGM}
\end{equation} where $\tau_{\rm IGM}(\nu)$ is the optical depth of the intervening IGM at frequency $\nu$ (computed as described in \S \ref{sec:IGM}), and $J(\nu)$ is the normalized (i.e. $\int J(\nu)d\nu=1$) Ly$\alpha$ spectrum of Ly$\alpha$ photons that emerges from the galaxy (computed as described in \S \ref{sec:winds}). The integral runs from $\sim 10^3$ km s$^{-1}$ blueward to $\sim 3\times 10^3$ km s$^{-1}$ redward of the Ly$\alpha$ line resonance, which spans the full range of velocities that is covered by the Ly$\alpha$ profile that emerges from the galaxy. The quantity $\mathcal{T}_{\rm IGM}$ is also referred to as the `IGM transmission fraction'. 
\begin{figure*}
\vbox{\centerline{\epsfig{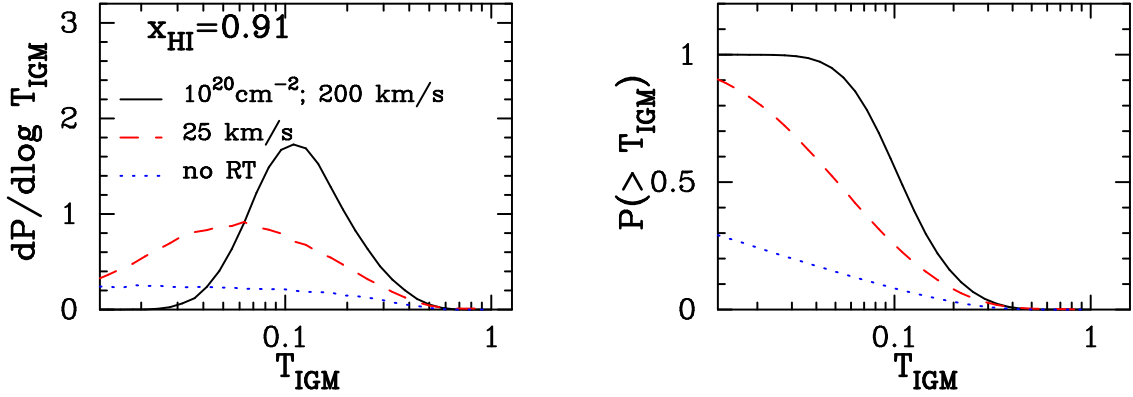}}}
\caption[]{{\it Left panel}: This Figure shows the probability density, $\frac{dP}{d\log \mathcal{T}_{\rm IGM}}$, that
  the IGM transmits to the observer a fraction of emitted Ly$\alpha$
  photons in the range $\log \mathcal{T}_{\rm IGM} \pm d \log \mathcal{T}_{\rm IGM}/2$, for
  galaxies populating dark matter halos of $10^{10}M_{
\odot} < M_{\rm halo} < 3\times 10^{10} M_{\odot}$ in a universe with a neutral fraction of $\avenf=0.91$
  (by volume) at $z=8.6$.  The {\it red dashed line} ({\it black solid
    line}) shows the model with $v_{\rm wind}=25$ (200) km s$^{-1}$. In both models $N_{\rm HI}=10^{20}$ cm$^{-2}$. The {\it blue
    dotted line} shows the `no-RT' model (see text). This Figure
  illustrates that ({\it i}) the IGM can transmit a significant
  fraction of  Ly$\alpha$ photons, despite the fact that reionization
  has only just started (text), and ({\it ii}) the IGM becomes even
  more `transparent' when winds are affecting Ly$\alpha$ scattering in
  the ISM. {\it Right panel}: Same as the {\it left panel}, but now we
  plot the cumulative distribution function (CDF), $P(>\Tigm)$. We find for example that
  $\mathcal{T}_{\rm IGM}> 10\%$  for $\sim 10\%$ of all halos in the `no-RT'
  model, and that this fraction is boosted when winds are present.}
\label{fig:pdf1}
\end{figure*}
\begin{figure*}
\vbox{\centerline{\epsfig{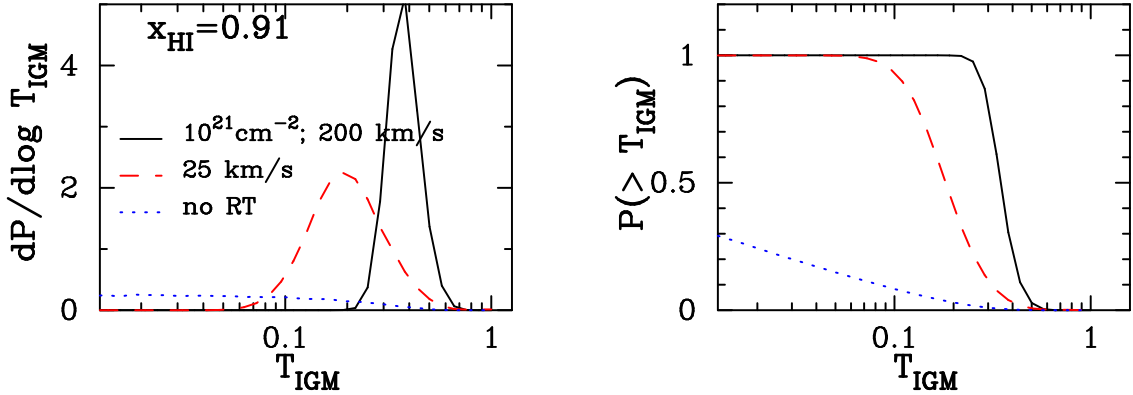}}}
\caption[]{Same as Figure~\ref{fig:pdf1}, but with an enhanced HI
  column density in the wind of $N_{\rm HI}=10^{21}$
  cm$^{-2}$. Frequency diffusion that occurs as Ly$\alpha$ photons
  scatter through an extremely opaque medium causes a larger fraction
  of photons to emerge at larger redshifts from the line line center,
  which reduces the effective IGM opacity.}
\label{fig:pdf2}
\end{figure*}

\subsection{An Example Line Profile}
\label{sec:line}

An example on an observed \lya\ line profile is shown in
Figure~\ref{fig:spec}. In this plot, the {\it solid line} shows the
Ly$\alpha$ spectrum-in units of the peak flux density-that emerges
from the galaxy after the photons have scattered through the HI
outflow (in this case $v_{\rm wind}=200$ km s$^{-1}$, and $N_{\rm HI}=10^{20}$ cm$^{-2}$). Most of the line flux is systematically redshifted relative
to the galaxy's systemic velocity. The flux density peaks at $\sim 2
v_{\rm wind}$, which is expected for radiation that scattered to the
observer from the outflowing gas on the far side of the Ly$\alpha$
source \citep[see][]{Ahn03,V06,DW10}.

The {\it blue dotted line} shows the IGM opacity, or more precisely $\exp-[\tau_{\rm IGM}(\nu)]$
as a function of velocity off-set $\Delta v$ for $\avenf=0.51$. In
this particular example infalling (ionized) gas provides a large
opacity to Ly$\alpha$ photons, even at velocities redward of the
Ly$\alpha$ line center \citep[in the frame of the galaxy, e.g.][]{Santos04,preIGM,IGM,Iliev08,Dayal10,ZZ1,Laursen11}. That is,
$\tau_{\rm IGM}\gg 1$ at $\Delta v \lsim +170$ km s$^{-1}$. At redder wavelengths there is no gas that falls in fast enough for the Ly$\alpha$
photons to appear at resonance. At these frequencies ($\Delta v \gsim
170$ km s$^{-1}$) the IGM opacity is dominated by the damping wing
optical depth of the neutral IGM which is a smooth function of
frequency.

The {\it red dashed line} shows the spectrum of photons after
processing the flux through the IGM, i.e. $J(\nu)\exp[-\tau_{\rm
    IGM}(\nu)]$. For this particular example, a fraction $\mathcal{T}_{\rm
  IGM}=0.57$ of all photons is transmitted directly to the
observer. For comparison, had we assumed that all photons emerged from
the galaxy with a Gaussian emission line, centered on the galaxy's systemic velocity and with a
standard deviation of $\sigma=v_{\rm circ}\sim 80$ km s$^{-1}$, we
would have found that $\mathcal{T}_{\rm IGM}=0.01$.
\begin{figure*}
\vbox{\centerline{\epsfig{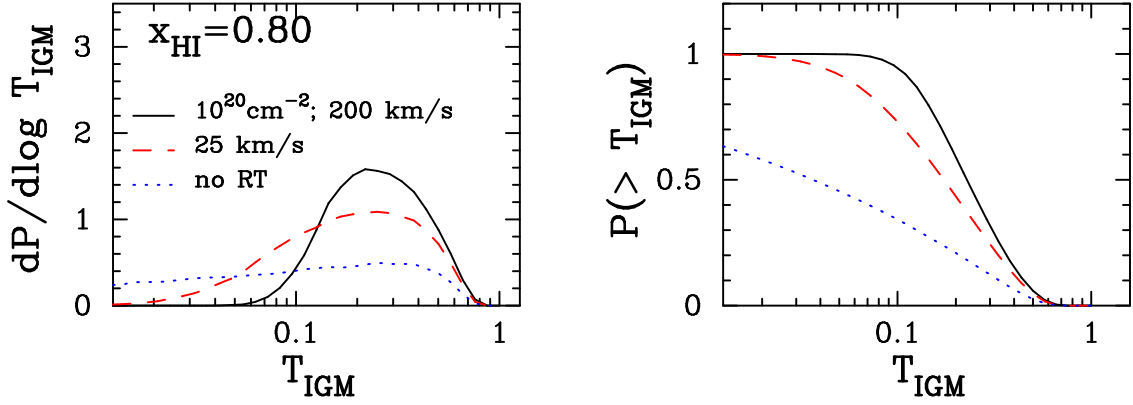}}}
\caption[]{Same as Figure~\ref{fig:pdf1}, but with a reduced global
  neutral hydrogen fraction, $\avenf=0.80$. The shifts of the $\mathcal{T}_{\rm
    IGM}$-PDFs are due to the
 reduced neutral hydrogen content of
  the Universe.}
\label{fig:pdf3}
\end{figure*}

\begin{figure*}
\vbox{\centerline{\epsfig{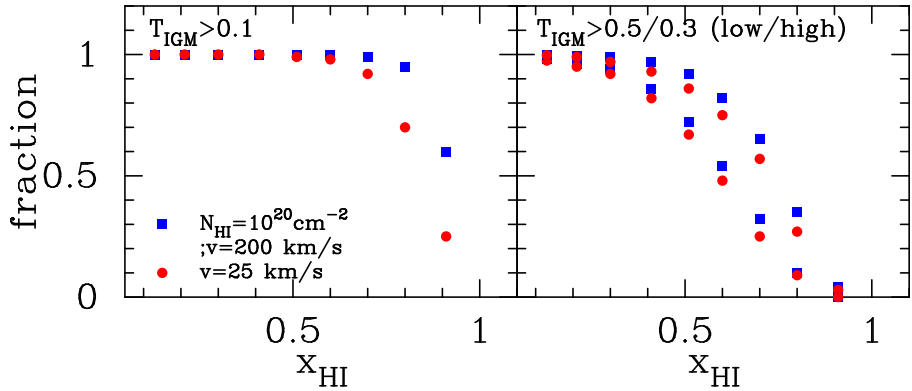}}}
\caption[]{{\it Left panel}: The fraction of halos with $10^{10}M_{
\odot} < M_{\rm halo} < 3\times 10^{10} M_{\odot}$ for which $\mathcal{T}_{\rm
    IGM}>0.1$ as a function of $\avenf$, assuming $v_{\rm wind}=25$
  (200) km s $^{-1}$, denoted by {\it red filled circles} ({\it blue
    filled squares}). In both models $N_{\rm HI}=10^{20}$
  cm$^{-2}$. {\it Right panel}: same as the {\it left panel}, but now
  we plot the fraction of halos for which $\mathcal{T}_{\rm IGM} >0.3$ (top set
  of points), and $\mathcal{T}_{\rm IGM} > 0.5$ (bottom set of points). This
  panel shows that the fraction of halos with $\mathcal{T}_{\rm IGM} >0.3$ only
  becomes  less that $\sim 10\%$ when $\avenf \gsim 0.90$.  We
  therefore conclude that the observed strength of the Ly$\alpha$ line
  in the $z=8.56$ galaxy is not surprising, unless the Universe was
  more than $90\%$ neutral.}
\label{fig:fvsz}
\end{figure*}

\subsection{The Probability Distribution Function (PDF) of $\mathcal{T}_{\rm IGM}$}
In Figure~\ref{fig:pdf1}, we show the PDF,  $\frac{dP}{d\log \mathcal{T}_{\rm IGM}}$ ({\it left panel}), and the
cumulative distribution function (CDF), $P(>\Tigm)$ ({\it right
  panel}), for $\avenf=0.91$.  The {\it red dashed lines} ({\it black
  solid lines}) correspond to models with $v_{\rm wind}=25$ (200) km
s$^{-1}$, and $N_{\rm HI}=10^{20}$ cm$^{-2}$ (both models). For the
low (high) wind velocity we find that the $\log \mathcal{T}_{\rm IGM}$--PDF
peaks around $\sim 0.06$ (0.1).  Increasing the wind velocity clearly
shifts the $\mathcal{T}_{\rm IGM}$--PDF to larger values. As the \lya\ photons
scatter off of the receeding outflows, they are Doppler shifted to
larger effective redshifts (e.g., Ahn et al. 2003, Verhamme et
al. 2006). Therefore, larger wind velocities cause a larger fraction
of Ly$\alpha$ photons to emerge at frequencies where they are immune
to the opacity in the ionized infalling IGM (see Fig.~\ref{fig:spec}).
Furthermore, this Doppler-shifting from the winds means that by the
time the \lya\ photons reach a neutral patch of the IGM, their
absorption cross-sections are further out on the damping wing
tail. Therefore, outflows reduce the impact of both resonant and
damping with opacities (see Fig~\ref{fig:spec}). Specifically, the right panel shows that $\mathcal{T}_{\rm
  IGM}> 10\%$  for $\sim 30\%$ ($\sim 60\%$) of LAEs for $v_{\rm
  wind}=25$ km s$^{-1}$ ($v_{\rm wind}=200$ km s$^{-1}$). 
 
To underline the effect of winds,  we compare to a model in which we only
evaluate the damping wing optical depth, $\tau_{\rm D}$ at  line
center, i.e. we set $\mathcal{T}_{\rm IGM} \equiv -\ln \tau_{\rm D}(\Delta \nu=0)$.
 This model is referred to as the `no-RT' model, as it
corresponds to a model in which no scattering of Ly$\alpha$ photons
occurs in the either the ISM or the ionized IGM, and is represented by
the {\it blue dotted line}. The $\mathcal{T}_{\rm IGM}$ is clearly skewed more
to lower values for this no-RT (`RT' stands for radiative transfer) model. That is, without winds a neutral
IGM dramatically attenuates the transmission of the \lya\ line. Only
$\sim$10\% of LAE have transmission fractions greater than 0.1  (as
has been demonstrated repeatedly in the past, e.g., Cen \& Haiman
2000, Santos 2004; Furlanetto et al. 2004; McQuinn et al. 2008;
Mesinger \& Furlanetto 2008a,b). Winds therefore clearly boost the
detectability of Ly$\alpha$ emission from galaxies during the early
phases in the EoR.

Figure~\ref{fig:pdf2} shows the same quantities as
Figure~\ref{fig:pdf1}, but for wind models with $N_{\rm HI}=10^{21}$
cm$^{-2}$. The `no-RT' model is of course unchanged. The $\mathcal{T}_{\rm
  IGM}$-PDFs are shifted to larger values, because resonant scattering
through very opaque media (the line-center optical depth to Ly$\alpha$
photons is $\tau_0=5.9\times 10^{7}(N_{\rm HI}/10^{21}\hs{\rm
  cm}^{-2})(T_{\rm gas}/10^4\hs{\rm K})^{-1/2}$), results in frequency
diffusion which increases with optical depth $\tau_0$ (e.g. Harrington 1973, Neufeld 1990). As a result of
this frequency diffusion, a larger fraction of photons will emerge at
larger redshifts from the line line center, which further reduces the
effective IGM opacity.

Figure~\ref{fig:pdf3} shows the same as Figure~\ref{fig:pdf1}, but for
a lower volume averaged neutral fraction of $\avenf=0.80$. We find
that $\mathcal{T}_{\rm IGM}> 10\%$ for $\sim 35\%$ for the no-RT model, and
that $\mathcal{T}_{\rm IGM}> 10\%$ for $\sim 75\%$ ($\sim 95\%$) of all halos
for the wind model with $v_{\rm wind}=25$ km s$^{-1}$ ($v_{\rm wind}=200$ km s$^{-1}$). This shift of the $\mathcal{T}_{\rm IGM}$-PDF arises
because of the reduced neutral hydrogen content of the Universe.
\section{Comparison to Recent Data}
\label{sec:discuss}

\subsection{Interpretation of the Recent Observations of a $z=8.56$ Galaxy}
\label{sec:z8.6}
The observed Ly$\alpha$ luminosity of the $z=8.56$ galaxy is
$L_{\alpha}=[5.5 \pm 1.0 \pm 1.8]\times 10^{42}$ erg s$^{-1}$ (the first number denotes the $1-\sigma$-uncertainty, while the second denotes the systematic uncertainty), while the UV luminosity density is $L_{\nu}(\lambda=1700\hs {\rm \AA})=10^{28.3\pm 0.2}$ erg s$^{-1}$ Hz$^{-1}$ at $\lambda=1700$ \AA\hs (rest-frame, Lehnert et al. 2010). From these observed
strengths of the line and continuum, we can constrain the observed
Ly$\alpha$ rest frame equivalent width (REW) to be \citep{Dwes10}:
\begin{eqnarray}
{\rm REW}=\frac{L_{\alpha}}{L_{\lambda}(\lambda=\lambda_{\alpha})}=\frac{\lambda_{\alpha}}{\nu_{\alpha}}\frac{L_{\alpha}}{L_{\nu}(\lambda=\lambda_{\alpha})}= \\ \nonumber
=\frac{\lambda_{\alpha}}{\nu_{\alpha}}\frac{L_{\alpha}}{L_{\nu}(\lambda=1700\hs\AA)}\Big{(}\frac{1216}{1700}\Big{)}^{\beta-2}=136^{+88}_{-55} \pm 45\hs{\rm \AA}. 
\end{eqnarray} 
We assumed that the UV continuum slope is $\beta = 2$,
which is appropriate for star forming galaxies with strong Ly$\alpha$
emission at $z=3-7$ \citep{Stark10}. \citet{Bouwens10b} determined $\beta\sim 3$ for candidate $z\sim 7$ ($z_{850}$ drop-out) galaxies, with $M_{\rm UV}\sim -19$ to $\sim -18$. Inserting $\beta=3$ into Eq~2 reduces the expectation value for REW to REW$=97$ \AA.

The quoted $1-\sigma$ uncertainty on REW is dominated by the uncertainty in the observed UV flux density.  The uncertainty on the observed REW is large, but the expectation value is remarkably large (even for $\beta=3$). The intrinsic Ly$\alpha$ REW, REW$_{\rm int}$, depends quite strongly on gas metallicity, and whether the galaxy is forming stars in a burst or continously \citep[e.g. Fig~7 of][]{S03}, and whether the `case-B' approximation is valid \citep{Raiter10}. The maximum possible value appears to be REW$_{\rm max}=3000$ \AA\hs \citep{Raiter10}. This maximum value is reached if this galaxy formed stars from metal-free gas in a burst with a top-heavy initial mass function. Under this assumption, the data requires that the IGM transmits $\mathcal{T}_{\rm IGM}\sim 0.04^{+0.03}_{-0.02}\times\Big{(}\frac{{\rm REW}_{\rm int}}{3000\hs{\rm \AA}}\Big{)}^{-1}$, where REW$_{\rm int}$ denotes the intrinsic (or emitted) REW. In theory, we could explain the observed REW for $\mathcal{T}_{\rm IGM}\gsim 0.04^{+0.03}_{-0.02}$. However, throughout we focus on the more conservative requirement $\mathcal{T}_{\rm IGM}>0.1$, which requires smaller values for the intrinsic REW. 

For comparison, Figures~\ref{fig:pdf1}-\ref{fig:pdf3} show that even
with low velocity outflows ($v_{\rm wind}=25$ km s$^{-1}$) $\mathcal{T}_{\rm
  IGM}>0.1$ for a significant fraction LAEs, even if the Universe is
$\sim 91\%$ neutral. This is illustrated more explicitly in
Figure~\ref{fig:fvsz}, where the {\it left panel} shows the fraction
of LAEs for which $\mathcal{T}_{\rm IGM}>0.1$ as a function of $\avenf$ for
$v_{\rm wind}=25$ (200) km s $^{-1}$ as {\it red filled circles} ({\it
  blue filled squares}), where $N_{\rm HI}=10^{20}$ cm$^{-2}$ in both
models.  This fraction is $\gsim 50\%$ for both models when $\avenf \lsim 80 \%$. The {\it right panel} is the same as the {\it left
  panel}, but for $\mathcal{T}_{\rm IGM}> 0.3$ (upper set of points), and
$\mathcal{T}_{\rm IGM}> 0.5$ (lower set of points). This panel shows that the
fraction of LAEs with $\mathcal{T}_{\rm IGM} >0.3$ only becomes less than $\sim
10\%$ when $\avenf \gsim 0.80$. We therefore conclude that the
strength of the Ly$\alpha$ line of the $=8.56$ galaxy is not
surprising, unless the Universe were more than $90\%$ neutral by
volume.  

\begin{figure}
\vbox{\centerline{\epsfig{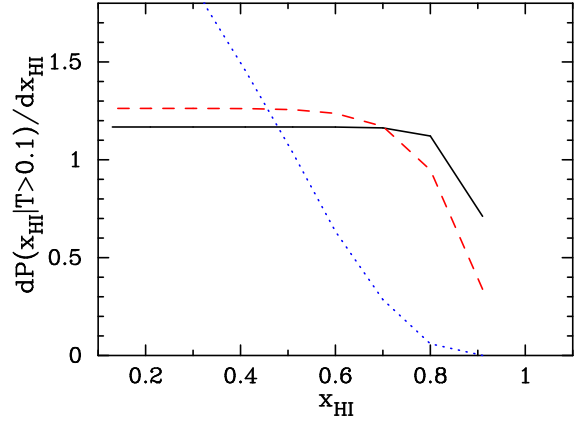}}}
\caption[]{This figure shows the marginalized PDF for $\avenf$, given that $\mathcal{T}_{\rm IGM}>0.1$, denoted by  $\frac{dP(\avenf | \mathcal{T}_{\rm IGM}>0.1)}{d \avenf}$. Different lines represent the different models that were also shown in Figures 2-4. This Figure shows clearly that it is difficult to rule out even large values of $\avenf$  with confidence--especially when winds are included in the modeling (see text). }
\label{fig:bayes}
\end{figure}
This latter statement is quantified in Figure~\ref{fig:bayes}, where we plot the conditional probability distribution for $\avenf$, {\it if we require that $\mathcal{T}_{\rm IGM}>0.1$} (the three lines represent the same models that were shown in Fig~2-4). This PDF follows from Bayes theorem as\footnote{Note that we switched to the shorter notation from probability theory, and denote probability density functions by $p(a)\equiv \frac{dP(a)}{d a}$, and conditional PDFs by  $p(a|b)\equiv \frac{dP(a|b)}{d a}$}  $p(\avenf | \mathcal{T}_{\rm IGM})=p(\mathcal{T}_{\rm IGM}| \avenf)p_{\rm prior}(\avenf)/p(\mathcal{T}_{\rm IGM})$.  We explicitly computed the term $\frac{dP(\mathcal{T}_{\rm IGM}| \avenf)}{d \log \mathcal{T}_{\rm IGM}}={\rm ln}\hs 10 \times \mathcal{T}_{\rm IGM}\times p(\mathcal{T}_{\rm IGM}| \avenf)$ in this paper. Examples were shown in Figure~2 ($\avenf=0.91$) and Figure~4 ($\avenf=0.80$). The term $p_{\rm prior}(\avenf)$ is the prior probability distribution for $\avenf$, which we assumed to be flat. This expression only gives us the conditional PDF for $\avenf$ for a given value of  $\mathcal{T}_{\rm IGM}$. To get a marginalized PDF, we compute $p(\avenf|\mathcal{T}_{\rm IGM}>0.1)=\int_{0.1}^{1.0} p(\avenf | \mathcal{T}_{\rm IGM}) p(\mathcal{T}_{\rm IGM})d\mathcal{T}_{\rm IGM}$, which simplifies to $p(\avenf|\mathcal{T}_{\rm IGM}>0.1)=\int_{0.1}^{1.0}  p(\mathcal{T}_{\rm IGM}| \avenf)d\mathcal{T}_{\rm IGM}$ (for our assumption that $p_{\rm prior}(\avenf)=1$).

If we require that $\mathcal{T}_{\rm IGM}>0.1$, then the {\it blue dotted line} shows that for the model without winds, a low neutral fraction is preferred. However, the statistical significance of this statement is weak: for this model $\avenf<0.70$ at $\sim 95\%$ CL. Once winds are included, this constraint becomes even weaker, and it is not even possible to `strongly' (i.e. $>95\%$ CL) rule out\footnote{If we require that $\mathcal{T}_{\rm IGM}>0.3$, then we can rule out $\avenf \geq 0.80$ at $\sim 95\%$ CL for both wind models. Given existing uncertainties on both observed and intrinsic REW of the Ly$\alpha$ line, requiring the IGM to transmit more than 30\% is not well motivated. We therefore consider any constraints on $\avenf$ from this requirement not relevant.} $\avenf \geq 0.91$. We caution against interpreting these numbers literally, as the overall shape of the PDF is completely dominated by the assumed prior on $\avenf$. This illustrates that it is difficult to rule out even large values of $\avenf$ with confidence--especially when winds are included in the modeling.

\subsection{Interpreting the Apparent Fast Drop in the `LAE Fraction' among LBGs}
\label{sec:z7}

\begin{figure}
\vbox{\centerline{\epsfig{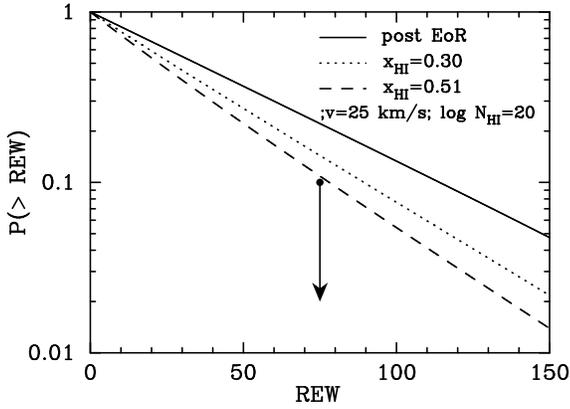}}}
\caption[]{We show the CDF for REW. The {\it solid line} shows an exponential Ly$\alpha$
  REW-distribution that represents a simplified representation of the
  $z=6$ drop-out population (see text). 
  The {\it dashed line} [{\it dotted line}] shows
  the $z=7$ PDF under the assumption that the observed LAE fraction at
  $z=7$ is different only because of the IGM, that the Universe
  was 51\% [30\%] neutral by volume (i.e. $\avenf=0.51$ [$\avenf=0.30$]), and for the
  wind model with $(N_{\rm HI},v_{\rm wind})=(10^{20}$cm$^{-1}$,25 km
  s$^{-1}$). We also show the upper limit that was derived by \citet{Stark10,Stark11} for $z=7$ drop-out galaxies. Explaining the observed rapid change in the LAE fraction
  among the drop-out population with reionization would 
require a fast evolution of the neutral fraction of hydrogen in the Universe. }
\label{fig:ewpdf}
\end{figure}

\citet{Stark10} recently found that the fraction of drop-out galaxies
with strong Ly$\alpha$ emission decreases strongly from $z=6$ to $z=7$
\citep[also see][]{Stark11}. More specificially, they found that the
fraction of drop-out galaxies with a  REW$\gsim 75$ \AA \hs \lya\ line
decreased by a factor of $\sim 2$ between $z=6$ and $z=7$. A similar
observation was made by \citet{Fontana10}, who detected (weak;
REW$\sim 13$ \AA) Ly$\alpha$ emission in only 1 out of 7 candidate
$z=7$ drop-out galaxies. These trends are not seen at $z\lsim6$.  Such
a sudden, strong suppression of Ly$\alpha$ flux from galaxies at $z=7$
could be a signature related to reionization, although the
available data still has large uncertainties. The \citet{Stark10,Stark11}
sample is consistent with no redshift evolution between $z=6$ and
$z=7$ at the $\sim $1 $\sigma$ level. And the statistical significance
of the drop observed by \citet{Fontana10} implicitly relies on the
assumption that all seven drop-out galaxies are indeed at $z\sim7$. Nevertheless, it is an interesting exercise to interprete these observations as is.

In this work, we computed the $\mathcal{T}_{\rm IGM}$-PDF at various ionization
stages of the IGM during the EoR. Suppose that the observed
REW-distribution at $z=6$ is described by an exponential function
(which provides a good fit to observed LAEs at $z=2-4$, see Gronwall
et al. 2007, Blanc et al. 2011) between REW$=0$ and REW$=300$ \AA, i.e. $P_{\rm z=6}({\rm
  REW})\propto$ exp [-REW/REW$_{\rm c}$]. If we choose a scalelength
of REW$_{\rm c}=50$ \AA, then the fraction of drop-out galaxies with
REW$\gsim 75$ \AA\hs is $f\sim 0.2$, which corresponds to the median
value observed by \citet{Stark10} at $z\sim6$.

We now conservatively assume that the IGM at $z=6$ was 100\%
transparent to Ly$\alpha$ photons emitted by galaxies, and that the
observed REW-PDF at $z=7$ is different only because of evolution of
the ionization state of the IGM. Under this assumption, an observed
REW at $z=7$ requires an intrinsic (i.e. emitted) equivalent width of
REW/$\mathcal{T}_{\rm IGM}$, and we can compute the observed REW-PDF at $z=7$ as
$P_{\rm z=7}({\rm REW})=\mathcal{N}\int_0^1 d\mathcal{T}_{\rm IGM}P(\mathcal{T}_{\rm
  IGM})P_{\rm z=6}({\rm REW}/\mathcal{T}_{\rm IGM})$. Here, $P(\mathcal{T}_{\rm IGM})$
denotes the $\mathcal{T}_{\rm IGM}$-PDF computed in this paper. The
equation sums over all possible $\mathcal{T}_{\rm IGM}$ with the proper
probabilities that a galaxy had a Ly$\alpha$ line with REW/$\mathcal{T}_{\rm
  IGM}$ {\it and} that the IGM transmitted a fraction $\mathcal{T}_{\rm IGM}$.
Finally, the factor $\mathcal{N}$ normalizes $P_{\rm
  z=7}({\rm REW})$.  
  
  Figure~\ref{fig:ewpdf} shows our model CDF of REW for $z=6$ drop-out galaxies as the {\it solid line}, and for $z=7$ drop out galaxies\footnote{Of course, we computed the $\mathcal{T}_{\rm IGM}$-PDF at $z=8.6$, and now apply our models to $z=7$ data. As a result, our
  model overestimates the Gunn-Peterson optical depth. We therefore
  overestimated the damping wing opacity of the neutral IGM. If we had
  included the proper damping wing optical depth (i.e. at the correct
  redshift), we would again have required an even larger $\avenf$
  to obtain the same total optical depth. Our current inferred rate of
  the evolution of $\avenf$ is therefore conservative.}
$\avenf=0.51$ ($\avenf=0.30$) as the {\it dashed line} ({\it dotted line}).
We used the wind model with $(N_{\rm HI},v_{\rm
  wind})=(10^{20}$cm$^{-2}$, 25 km s$^{-1}$).  
  
\citet{Stark10,Stark11} put the upper limit on the fraction of drop-out galaxies with Ly$\alpha$ REW$\geq 75$ \AA\hs at $\lsim 0.10$ (also shown in Fig~7). The plots show that at in order to explain the observed evolution between $z=6$ and $z=7$, we
require a rapid evolution of the neutral fraction of hydrogen in the
Universe (i.e. $\Delta \avenf \sim 0.5$ over $\Delta
z=1$)\footnote{Our constraint on the redshift-evolution in $\avenf$
  depends somewhat on the assumed range over which exponential
  function provides a good fit to the data. For example, under the
  extreme assumption that there were no drop-out galaxies at $z=6$
  with REW$\leq 20$ \AA, then we would need a scalelength of REW$_{\rm
    c}\sim 40$ \AA\hs to be consistent with \citet{Stark10}. In
  this case, we would `only' require that $\Delta \avenf \sim 0.3$
  over $\Delta z=1$. However \citet{Stanway07} found that the fraction
  of drop-out galaxies with weak emission (REW$\lsim 25$ \AA) is
  consistent with the observed fraction at $z=3$, which suggests that
  this `extreme' model is unrealistic.}.  We stress that we assumed
that $\mathcal{T}_{\rm IGM}=1$ at $z=6$. If we had instead used a $\mathcal{T}_{\rm
  IGM}$-PDF appropriate for a fully ionized medium at $z=6$, then we
would have needed the IGM at $z=7$ to be even more opaque, which
required an even faster evolution in $\avenf$. As we mentioned already
in \S~\ref{sec:intro}, dust (as well as evolution in metallicity)
would also require a faster evolution in $\avenf$.  For larger wind
velocities and/or HI column densities, we would again need a larger
volume fraction of HI. Additionally, the data of \citet{Fontana10}
implies an even stronger evolution in the observed REW-PDF, and hence
in the overall neutral fraction. 

Theoretically, the above-inferred rapid redshift evolution in $\avenf$
is unrealistic even in models with no negative feedback on the source
population (e.g., Barkana \& Loeb 2001; Fig. 9 in Mesinger et
al. 2006; Fig. 1 in Lidz et al. 2007).  Furthermore, the sinks of
ionizing photons (Lyman limit absorption systems) likely further slow
the final stages of reionization (e.g., Furlanetto \& Mesinger 2009;
Alvarez \& Abel 2010; Crociani et al. 2011), whose photoevaporation
timescales could be much longer than $\Delta z\sim1$ (Iliev et
al. 2005). The inferred rapid evolution could mean that the current
sample of $z=7$ drop-out galaxies trace a region of the Universe that
was more neutral than average. Alternatively, it could signal other
physical effects: for example, a decreasing wind strength or covering factor towards
higher redshifts can enhance the impact of the ionized IGM, and thus
increase the rate at which the IGM opacity changes. Yet another
interesting possibility is that the observed rapid evolution in the
Ly$\alpha$ REW-PDF of the drop-out galaxy population is entirely due
to reionization, but that the Universe at $z=6$ still contained a
non-negligible volume fraction of neutral hydrogen (for a discussion
on the current observational constraints, see Mesinger 2010 and
McGreer et al., 2011).
 This is because the fraction of galaxies for which $\mathcal{T}_{\rm
   IGM}>0.3$ (as an example) evolves more rapidly at $\avenf \gsim 0.5$
 (i.e. $df/d\avenf$ is largest when $\avenf\sim 0.5$; see Fig~\ref{fig:fvsz}). 
 For example, the fraction of galaxies with $\mathcal{T}_{\rm IGM}>0.3$ changes more rapidly
 between $\avenf =0.5$ and $\avenf =0.7$, then between $\avenf =0.0$ and $\avenf =0.5$ (see Fig~\ref{fig:fvsz}). 
 This scenario however
 would require an extended early epoch of reionization to be
 consistent with WMAP observations, perhaps driven by negative
 feedback on smaller-mass sources.  Clearly it will be important to
 constrain the evolution of the LAE fraction between $z=6$ and $z=7$
 with a larger sample of galaxies.

Finally, we point out that the observed drop in the LAE fraction
between $z=6$ and $z=7$ is consistent with observations of the LAE
populations at $z=5.7$ and $z=6.5$. The Ly$\alpha$ luminosity function
of LAEs evolves significantly between $z=6.5$ and $z=5.7$
\citep[][]{Shima06,Ka06,Ouchi10}. However, the rest-frame UV
luminosity function of these same galaxies does not evolve between
these redshifts (Kashikawa et al. 2006). \citet{LF} showed that
these two observations combined translate to a reduction in the number
of detected Ly$\alpha$ photons from $z=6.5$ by a factor of $1.1-1.8$
(95\% CL) relative to $z=5.7$, with a median value of $\sim 1.3$
\citep[][]{Ouchi10}. Hu et al. (2010) recently obtained spectroscopic
observations of narrowband selected LAEs at $z=5.7$ and $z=6.5$ and
found that the Ly$\alpha$ REW of the average $z=5.7$ [$z=6.5$]
spectrum was REW$\sim 34\pm 2$ \AA\hs [REW$\sim 23\pm 3$ \AA]. This
corresponds to a reduction in the number of detected Ly$\alpha$
photons from $z=6.5$ by a factor of $\sim 1.3$ relative to $z=5.7$,
consistent with the value inferred from the redshift evolution of the
luminosity functions. This provides evidence for evolution
in the REW-PDF that is similar to the trends seen in the drop-out population by \citet{Stark10}
and \citet{Fontana10}.

\section{Comparison to Previous Work}
\label{sec:prevwork}

We found that  $\mathcal{T}_{\rm IGM}\gsim 50\%$ for the majority of galaxies, even when the Universe is $\sim 60 \%$ neutral by volume. In our models, the IGM is more transparent than in previous studies of the visibility of LAEs during the EoR. Mesinger \& Furlanetto (2008b) assumed a constant opacity in the ionized IGM, and only considered velocity offsets in the Ly$\alpha$ line due to the peculiar velocities of the dark matter halos (which are much smaller than the velocity offsets that can be imparted by winds). McQuinn et al. (2007) studied a very similar model, but did not include the halos' peculiar velocities. McQuinn et al. (2007) also investigated a `wind model' in which the Ly$\alpha$ line was redshifted by $400$ km s$^{-1}$, and found a considerable boost in the IGM transmission fraction for $\avenf \gsim 0.7$ (see their Figure~6). Iliev et al. (2008) studied the impact of the ionized IGM on the Ly$\alpha$ line profile in more detail, but assumed Gaussian emission lines, which can reduce the IGM transmission factor significantly compared to models that include outflows (see \S~\ref{sec:line} for a clear illustration of this effect).

Perhaps surprisingly, our quoted values for $T_{\rm IGM}$ are significantly higher than the values quoted in previous works {\it even for a fully ionized IGM}. For example, \citet{IGM} found that the ionized IGM could transmit as little as $\mathcal{T}_{\rm IGM}$=0.1-0.3 at $z>4.5$ for a fully ionized IGM (also see Dayal et al. 2010). Laursen et al. (2011) recently constrained $\mathcal{T}_{\rm IGM}=0.26^{+0.13}_{-0.18}$ at $z=5.7$. \citet{ZZ1} find that $\mathcal{T}_{\rm IGM} \sim 0.01-0.3$, which in detail depends on luminosity\footnote{Note that \citet{ZZ1} include scattering inside the virial radius in their calculations, while for example \citet{IGM} and \citet{Laursen11} do not. Gas within the virial radius can be significantly denser and can have larger peculiar velocities. This can explain that \citet{ZZ1} found lower values for $\mathcal{T}_{\rm IGM}$. The fact that \citet{ZZ1} properly account for the possibility that photons scatter back into the LOS (in close proximity of the source) can only boost their $\mathcal{T}_{\rm IGM}$ compared to that of other groups, and cannot explain this difference.}.

The main difference between the present and previous work are due to outflows, which causes most of the line flux to be systematically redshifted relative to the galaxy's systemic velocity (see \S~\ref{sec:result}). We acknowledge that the outflow models were calibrated on the models of Verhamme et al. (2008) which assumed that the IGM at $z=2-5$ had no impact on the observed Ly$\alpha$ line shapes. It is certainly possible--especially towards higher redshifts--to reproduce observed Ly$\alpha$ line shapes with `weaker' (i.e. lower velocity) outflows once the IGM is included. This is illustrated by recent work of \citet{Laursen11}, who modeled Ly$\alpha$ RT in simulated, dusty galaxies. \citet{Laursen11} found that Ly$\alpha$ photons emerge from their simulated galaxies with a broadened, double peaked profile (also see Barnes et al. 2011). The IGM at $z>5$ cuts off the blue peak, which then results in a typically observed redshifted, asymmetric Ly$\alpha$ emission line (although at $z=3.5$ the blue peak remains visible for a significant fraction of sightlines through the IGM). In these models, galactic outflows have little impact on the Ly$\alpha$ photons emerging from galaxies. 

While it is likely that Ly$\alpha$ line shape is affected by the IGM to some extend (especially towards higher redshifts), outflows are also expected to play at least an important role. This is because outflows are ubiquitous in observed star forming galaxies, and the outflowing material has a large covering factor \citep{Steidel10}. Importantly, there is evidence that outflows affect observed Ly$\alpha$ line spectra:  ({\it i}) asymmetric line shapes are present at low redshift ($z\sim 0$), when the IGM should not have an impact \citep{Mas03,H11}; ({\it ii}) observations indicate that outflows promote the escape of Ly$\alpha$ from a dusty medium \citep{Kunth98,Atek08}; ({\it iii}) \citet{SV08} and \citet{Des10} successfully reproduced the Ly$\alpha $ spectrum of the Lyman Break galaxies cB58 and the `8 o'clock arc' respectively, with outflow models very similar to our own and those by \citet{V06,V08}, but whose parameters were constrained by low-ionization metal absorption lines. In these models, the wind parameters inferred from the Ly$\alpha$ line shape were consistent with those inferred from alternative observations. Steidel et al. (2010,2011) also constructed a simple model for the Ly$\alpha$ spectra observed from LBGs--as well as the Ly$\alpha$ halos that are observed around LBGs--in which Ly$\alpha$ photons scatter primarily through a large-scale galactic outflow, whose structure is constrained by low-ionization metal absorption lines. Points ({\it i}--{\it iii}) underline the probable physical connection between the outflowing and scattering media. 

We have shown that our main conclusions remain valid for wind velocities in excess\footnote{If we lower $N_{\rm HI}$ by an order of magnitude, then our main conclusions are unaffected provided that $v_{\rm wind}\gsim 200$ km s$^{-1}$. These requirements are also reasonable.} of $v_{\rm wind} \gsim 25$ km s$^{-1}$ (and $N_{\rm HI} \geq 10^{20}$ cm$^{-2}$), which systematically redshifts the Ly$\alpha$ emission line as it emerges from a galaxy. We consider these requirements to be reasonable, given direct observational constraints on outflow velocities from low-ionization metal absorption lines \citep{Steidel10,Rakic10}, and on HI column density (as in cB58 and the '8 o'clock arc', see e.g. Schaerer \& Verhamme 2008 and Dessauges-Zavadsky et al. 2010),.

After the submission of this work, Dayal \& Ferrara (2011, hereafter DF11) submitted a paper in which they constrained $\avenf <0.2$ using the same $z=8.6$ galaxy. We argue that this upper limit is not robust for several reasons. 
Firstly, the model of DF11 does not include outflows. In contrast, we have summarized the strong observational evidence that galactic outflows affect the Ly$\alpha$ radiation field, implying that outflows must be included at some level in a realistic model. Secondly, there is an absence of large HII `bubbles' in the model of DF11: the HII bubble radii in their model are $\lsim 3-4$ Mpc when $\avenf=0.2$, while simulations of reionization that include radiative transfer of ionizing photons, show that the typical HII bubbles have radii that are a factor $\sim 10$ larger when $\avenf \sim 0.3$ (see e.g. Fig~2 of Zahn et al. 2010). Our large scale semi-numeric simulations of reionization properly capture the bubble size distribution, and the fact that the more massive galaxies preferentially populate these large HII bubbles. This absence of large bubbles in the models of DF11 enhances the IGM opacity, which causes their constraints on $\avenf$ to be stronger than those obtained from our models where galactic winds are not included.

\section{Conclusions}
\label{sec:conc}

In this paper we have studied the visibility of the Ly$\alpha$ emission line during
reionization.  We combine large scale semi-numerical simulations of
cosmic reionization with empirically-calibrated models of galactic
outflows. With these sophisticated tools, we compute the PDFs of the IGM
transmission fraction, $\mathcal{T}_{\rm IGM}$.
 We find that winds cause $\mathcal{T}_{\rm IGM}\gsim 10 \%$ [50\%],
  for the majority of galaxies, even when the Universe is $\sim 80 \%$
  [60\%] neutral by volume.
This only requires wind speeds greater than $\sim 25$ km  s$^{-1}$,
which are quite conservative judging by the observed \lya\ lines
shapes at $z<5$ (Verhamme et al. 2008, also see \S~\ref{sec:prevwork}).
Therefore, we conclude that the observed strong Ly$\alpha$ emission
from the reported $z=8.6$ galaxy is consistent with a highly neutral
IGM.

 We also show that evoking reionization to explain the observed drop
 in the `LAE fraction'  (see \S~\ref{sec:z7}) of drop-out galaxies
 between $z=6$ and $z=7$ \citep{Stark10,Stark11}, requires a very
 rapid evolution of $\mathcal{T}_{\rm IGM}$, corresponding to $\avenf \sim 0
 \rightarrow 0.5$ over $\Delta z=1$.  Reionization models find such a
 rapid evolution unrealistic, which may indicate that either ({\it i})
 the current sample of drop-out galaxies at $z=7$ happened to populate
 a region of our Universe that was more neutral than average, ({\it
   ii}) winds become weaker and/or have smaller covering factors towards higher redshifts, or ({\it iii}) that the Universe at $z=6$ still contained a non-negligible volume fraction of neutral hydrogen.  However, these conclusions are tentative as the available data still has large uncertainties. 

Regardless of these current observational uncertainties,
 our work underlines the point that Ly$\alpha$ emission can be detected from
 galaxies in the earliest stages of reionization. This is a positive result for (narrowband) searches for high redshift Ly$\alpha$ emitters such as the `Emission-Line galaxies with VISTA Survey' (ELVIS) \citep[e.g.][]{N07}. On the other hand, if a neutral IGM is quite transparent to Ly$\alpha$ photons, then a signature of reionization may be more difficult to extract from
 observations of Ly$\alpha$ emitting galaxies.
 However, the {\it redshift evolution} of quantities such as ({\it i})
 the `LAE fraction'--or more generally the Ly$\alpha$ restframe
 equivalent width PDF-- among LBGs \citep{Stark10,Fontana10,Stark11}, and ({\it ii}) the UV and
 Ly$\alpha$ luminosity functions of LAEs (Kashikawa et al. 2006), already provide interesting
 and useful constraints on models of reionization. Furthermore, the
 clustering signature of LAEs 
 (\citealt{McQuinn07}; Mesinger \& Furlanetto 2008b, though see Iliev et al. 2008) is also affected by reionization, and it has already been shown that
 winds do not affect this prediction \citep{McQuinn07}.

\vskip+0.5in
{\bf Acknowledgements}
Support for this work was provided by NASA through Hubble Fellowship
grant HST-HF-51245.01-A to AM, awarded by the Space Telescope Science
Institute, which is operated by the Association of Universities for
Research in Astronomy, Inc., for NASA, under contract NAS 5-26555.


\label{lastpage}

\begin{thebibliography}{14}
\expandafter\ifx\csname natexlab\endcsname\relax\def\natexlab#1{#1}\fi
\bibitem[Ahn et al.(2003)]{Ahn03} Ahn, S.-H., Lee, H.-W.,  \& Lee,
H.~M.\ 2003, \mnras, 340, 863

\bibitem[Alvarez 
\& Abel(2010)]{Alvarez10} Alvarez, M.~A., \& Abel, T.\ 2010, arXiv:1003.6132 

\bibitem[Atek et 
al.(2008)]{Atek08} Atek, H., Kunth, D., Hayes, M., {\"O}stlin, G., \& Mas-Hesse, J.~M.\ 2008, \aap, 488, 491 

\bibitem[Atek et 
al.(2009)]{Atek09} Atek, H., Schaerer, D., \& Kunth, D.\ 2009, \aap, 502, 791 

\bibitem[Barkana 
\& Loeb(2001)]{BL01} Barkana, R., \& Loeb, A.\ 2001, \physrep, 349, 125 

\bibitem[Barnes et al.(2011)]{Barnes11} Barnes, L.~A., Haehnelt, 
M.~G., Tescari, E., \& Viel, M.\ 2011, arXiv:1101.3319 

\bibitem[Blanc et al.(2011)]{Blanc10} Blanc, G.~A., et al.\ 
2011, submitted to ApJ, arXiv:1011.0430 

\bibitem[Bolton 
\& Haehnelt(2007)]{BH07} Bolton, J.~S., \& Haehnelt, M.~G.\ 2007, \mnras, 382, 325 

\bibitem[Bouwens et al.(2010a)]{Bouwens10} Bouwens, R.~J., et al.\ 
2010a, \apjl, 709, L133 

\bibitem[Bouwens et al.(2010b)]{Bouwens10b} Bouwens, R.~J., et al.\ 
2010b, \apjl, 708, L69 

\bibitem[Bunker et al.(2010)]{Bunker10} Bunker, A.~J., et al.\ 
2010, \mnras, 409, 855 

\bibitem[Cen 
\& Haiman(2000)]{Cen00} Cen, R., \& Haiman, Z.\ 2000, \apjl, 542, L75 

\bibitem[Cen et al.(2005)]{2005ApJ...621...89C} Cen, R., Haiman, Z., 
\& Mesinger, A.\ 2005, \apj, 621, 89 

\bibitem[Crociani et al.(2011)]{Croc10} Crociani, D., 
Mesinger, A., Moscardini, L., \& Furlanetto, S.\ 2011, \mnras, 411, 289 

\bibitem[Dayal et al.(2011)]{Dayal10} Dayal, P., Maselli, A., 
\& Ferrara, A.\ 2011, \mnras, 410, 830 

\bibitem[Dayal 
\& Ferrara(2011)]{Dayal} Dayal, P., \& Ferrara, A.\ 2011, arXiv:1102.1726 

\bibitem[Dessauges-Zavadsky et 
al.(2010)]{Des10} Dessauges-Zavadsky, M., D'Odorico, S., Schaerer, D., Modigliani, A., Tapken, C., \& Vernet, J.\ 2010, \aap, 510, A26 

\bibitem[Dijkstra et al.(2006a)]{D06} Dijkstra, M., Haiman, 
Z., \& Spaans, M.\ 2006a, \apj, 649, 14 

\bibitem[Dijkstra et al.(2006b)]{preIGM} Dijkstra, M., Haiman, 
Z., \& Spaans, M.\ 2006b, \apj, 649, 37 

\bibitem[Dijkstra et al.(2007a)]{LF} Dijkstra, M., Wyithe,  J.~S.~B.,
\& Haiman, Z.\ 2007a, \mnras, 379, 253

\bibitem[Dijkstra et al.(2007b)]{IGM} Dijkstra, M., Lidz,  A., \&
Wyithe, J.~S.~B.\ 2007b, \mnras, 377, 1175

\bibitem[Dijkstra \& Wyithe(2010)]{DW10} Dijkstra, M., \& Wyithe, J.~S.~B.\ 2010, \mnras, 408, 352 

\bibitem[Dijkstra 
\& Westra(2010)]{Dwes10} Dijkstra, M., \& Westra, E.\ 2010, \mnras, 401, 2343 

\bibitem[Fan et al.(2006)]{Fan06} Fan, X., et al.\ 2006, \aj, 
132, 117 

\bibitem[Finkelstein et al.(2010)]{Finkelstein10} Finkelstein, S.~L., 
Papovich, C., Giavalisco, M., Reddy, N.~A., Ferguson, H.~C., Koekemoer, 
A.~M., \& Dickinson, M.\ 2010, \apj, 719, 1250 

\bibitem[Fontana et al.(2010)]{Fontana10} Fontana, A., et al.\ 
2010, \apjl, 725, L205 

\bibitem[Furlanetto et al.(2004)]{Fur04} Furlanetto, S.~R., 
Zaldarriaga, M., \& Hernquist, L.\ 2004, \apj, 613, 1 

\bibitem[Furlanetto 
\& Mesinger(2009)]{2009MNRAS.394.1667F} Furlanetto, S.~R., \& Mesinger, A.\ 2009, \mnras, 394, 1667 

\bibitem[Gronwall et al.(2007)]{2007ApJ...667...79G} Gronwall, C., et al.\ 
2007, \apj, 667, 79 

\bibitem[Harrington(1973)]{Harrington73} Harrington, J.~P.\ 1973,
\mnras, 162, 43

\bibitem[Hayes et al.(2011)]{Hayes10} Hayes, M., Schaerer, D., 
{\"O}stlin, G., Mas-Hesse, J.~M., Atek, H., 
\& Kunth, D.\ 2011, \apj, 730, 8 

\bibitem[Heckman et al.(2011)]{H11} Heckman, T.~M., et al.\ 
2011, \apj, 730, 5 

\bibitem[Hu et al.(2010)]{Hu10} Hu, E.~M., Cowie, L.~L., 
Barger, A.~J., Capak, P., Kakazu, Y., \& Trouille, L.\ 2010, \apj, 725, 394 

\bibitem[Iliev et al.(2005)]{Iliev05} Iliev, I.~T., Shapiro, 
P.~R., \& Raga, A.~C.\ 2005, \mnras, 361, 405 

\bibitem[Iliev et al.(2008)]{Iliev08} Iliev, I.~T., Shapiro,  P.~R.,
McDonald, P., Mellema, G., \& Pen, U.-L.\ 2008, \mnras, 391, 63

\bibitem[Johnson et al.(2009)]{J09b} Johnson, J.~L., Greif,  T.~H.,
Bromm, V., Klessen, R.~S., \& Ippolito, J.\ 2009, \mnras, 399, 37

\bibitem[Kashikawa et al.(2006)]{Ka06} Kashikawa, N., et  al.\ 2006,
\apj, 648, 7

\bibitem[Komatsu et al.(2009)]{Komatsu08} Komatsu, E., et al.\ 
2009, \apjs, 180, 330 

\bibitem[Kunth et 
al.(1998)]{Kunth98} Kunth, D., Mas-Hesse, J.~M., Terlevich, E., Terlevich, R., Lequeux, J., \& Fall, S.~M.\ 1998, \aap, 334, 11 

\bibitem[Laursen et al.(2011)]{Laursen11} Laursen, P., 
Sommer-Larsen, J., \& Razoumov, A.~O.\ 2011, \apj, 728, 52 

\bibitem[Lehnert et al.(2010)]{Lehnert} Lehnert, M.~D., et al.\ 
2010, \nat, 467, 940 

\bibitem[Lidz et al.(2007)]{Lidz07} Lidz, A., McQuinn, M., 
Zaldarriaga, M., Hernquist, L., \& Dutta, S.\ 2007, \apj, 670, 39 

\bibitem[Mas-Hesse et al.(2003)]{Mas03} Mas-Hesse, J.~M., 
Kunth, D., Tenorio-Tagle, G., Leitherer, C., Terlevich, R.~J., 
\& Terlevich, E.\ 2003, \apj, 598, 858 

\bibitem[McGreer et al.(2011)]{McGreer11} McGreer, I.~D., 
Mesinger, A., \& Fan, X.\ 2011, submitted to \mnras, arXiv:1101.3314 

\bibitem[McQuinn et al.(2007)]{McQuinn07} McQuinn, M., Hernquist, 
L., Zaldarriaga, M., \& Dutta, S.\ 2007, \mnras, 381, 75 

\bibitem[McQuinn et al.(2008)]{McQuinn08} McQuinn, M., Lidz, A., 
Zaldarriaga, M., Hernquist, L., \& Dutta, S.\ 2008, \mnras, 388, 1101 

\bibitem[Mesinger et al.(2006)]{Mesinger06} Mesinger, A., Johnson, 
B.~D., \& Haiman, Z.\ 2006, \apj, 637, 80 

\bibitem[Mesinger 
\& Furlanetto(2007)]{MF07} Mesinger, A., \& Furlanetto, S.\ 2007, \apj, 669, 663 

\bibitem[Mesinger 
\& Furlanetto(2008a)]{Mes1} Mesinger, A., \& Furlanetto, S.~R.\ 2008a, \mnras, 385, 1348 

\bibitem[Mesinger  \& Furlanetto(2008b)]{Mes2} Mesinger, A., \&
Furlanetto, S.~R.\ 2008b, \mnras, 386, 1990

\bibitem[Mesinger 
\& Furlanetto(2009)]{Mesinger09} Mesinger, A., \& Furlanetto, S.\ 2009, \mnras, 400, 1461 

\bibitem[Mesinger(2010)]{Mesinger10} Mesinger, A.\ 2010, \mnras, 
407, 1328 

\bibitem[Mesinger et al.(2011)]{21cm} Mesinger, A., 
Furlanetto, S., \& Cen, R.\ 2011, \mnras, 411, 955 

\bibitem[Miralda-Escude(1998)]{M98} Miralda-Escude, J.\ 
1998, \apj, 501, 15 

\bibitem[Neufeld(1990)]{Neufeld90} Neufeld, D.~A.\ 1990, \apj,  350,
216

\bibitem[Nilsson et 
al.(2007)]{N07} Nilsson, K.~K., Orsi, A., Lacey, C.~G., Baugh, C.~M., \& Thommes, E.\ 2007, \aap, 474, 385 

\bibitem[Ouchi et al.(2010)]{Ouchi10} Ouchi, M., et al.\ 2010, 
\apj, 723, 869 

\bibitem[Partridge  \& Peebles(1967)]{1967ApJ...147..868P} Partridge,
R.~B., \& Peebles, P.~J.~E.\ 1967, \apj, 147, 868

\bibitem[Pawlik et al.(2011)]{Pawlik} Pawlik, A.~H., 
Milosavljevi{\'c}, M., \& Bromm, V.\ 2011, \apj, 731, 54 

\bibitem[Raiter et 
al.(2010)]{Raiter10} Raiter, A., Schaerer, D., \& Fosbury, R.~A.~E.\ 2010, \aap, 523, A64 

\bibitem[Rakic et al.(2010)]{Rakic10} Rakic, O., Schaye, J., 
Steidel, C.~C., \& Rudie, G.~C.\ 2010, arXiv:1011.1282 

\bibitem[Rybicki 
\& Lightman(1979)]{1979rpa..book.....R} Rybicki, G.~B., \& Lightman, A.~P.\ 1979, New York, Wiley-Interscience, 1979.~393

 
\bibitem[Santos(2004)]{Santos04} Santos, M.~R.\ 2004, \mnras, 
349, 1137 

\bibitem[Schaerer(2002)]{S02} Schaerer, D.\ 2002, \aap, 382, 28

\bibitem[Schaerer(2003)]{S03} Schaerer, D.\ 2003, \aap,  397, 527

\bibitem[Schaerer 
\& Verhamme(2008)]{SV08} Schaerer, D., \& Verhamme, A.\ 2008, \aap, 480, 369 

\bibitem[Shimasaku et al.(2006)]{Shima06} Shimasaku, K., et  al.\
2006, \pasj, 58, 313

\bibitem[Stanway et al.(2005)]{Stanway05} Stanway, E.~R., 
McMahon, R.~G., \& Bunker, A.~J.\ 2005, \mnras, 359, 1184 

\bibitem[Stanway et al.(2007)]{Stanway07} Stanway, E.~R., et al.\ 
2007, \mnras, 376, 727 

\bibitem[Stark et al.(2010a)]{Stark10} Stark, D.~P., Ellis, 
R.~S., Chiu, K., Ouchi, M., \& Bunker, A.\ 2010a, \mnras, 408, 1628 

\bibitem[Stark et al.(2011)]{Stark11} Stark, D.~P., Ellis, 
R.~S., \& Ouchi, M.\ 2011, \apjl, 728, L2 


\bibitem[Steidel et al.(2010)]{Steidel10} Steidel, C.~C., Erb, 
D.~K., Shapley, A.~E., Pettini, M., Reddy, N., Bogosavljevi{\'c}, M., 
Rudie, G.~C., \& Rakic, O.\ 2010, \apj, 717, 289 

\bibitem[Trac 
\& Cen(2007)]{TC07} Trac, H., \& Cen, R.\ 2007, \apj, 671, 1 

\bibitem[Vanzella et 
al.(2010)]{V10} Vanzella, E., et al.\ 2010, \aap, 513, A20 

\bibitem[Verhamme et al.(2006)]{V06} Verhamme, A.,  Schaerer, D., \&
Maselli, A.\ 2006, \aap, 460, 397

\bibitem[Verhamme et  al.(2008)]{V08} Verhamme, A., Schaerer, D.,
Atek, H., \& Tapken, C.\ 2008, \aap, 491, 89

\bibitem[Yan et al.(2010)]{Yan10} Yan, H.-J., Windhorst, 
R.~A., Hathi, N.~P., Cohen, S.~H., Ryan, R.~E., O'Connell, R.~W., 
\& McCarthy, P.~J.\ 2010, Research in Astronomy and Astrophysics, 10, 867 

\bibitem[Zahn et al.(2010)]{Zahn10} Zahn, O., Mesinger, A., 
McQuinn, M., Trac, H., Cen, R., \& Hernquist, L.~E.\ 2010, arXiv:1003.3455 

\bibitem[Zel'Dovich(1970)]{Z70} Zel'Dovich, Y.~B.\ 1970, \aap, 5, 84 

\bibitem[Zheng et al.(2010a)]{ZZ1} Zheng, Z., Cen, R., Trac, 
H., \& Miralda-Escud{\'e}, J.\ 2010a, \apj, 716, 574 

\bibitem[Zheng et al.(2010b)]{ZZ2} Zheng, Z., Cen, R., Trac, H., \&
Miralda-Escude, J.\ 2010b, submitted to ApJ, arXiv:1010.3017


\end{thebibliography}
\end{document}